\documentstyle[epsf,twocolumn,prl,aps,amstex]{revtex}

\begin{document}

\tighten                                                 

\draft
%

\wideabs{

\title
{
Neutron polarizabilities investigated by quasi-free Compton scattering from
the deuteron
}

\author
{
K.~Kossert$^1$, M.~Camen$^1$, F.~Wissmann$^1$, J.~Ahrens$^2$, H.-J.~Arends$^2$,
R.~Beck$^2$, G.~Caselotti$^2$, P.~Grabmayr$^3$, P.D.~Harty$^4$, O.~Jahn$^2$,
P.~Jennewein$^2$,
M.I.~Levchuk$^5$, A.I.~L'vov$^6$, J.C.~McGeorge$^4$,
A.~Natter$^3$, V.~Olmos~de~Le\'on$^2$, V.A.~Petrun'kin$^6$, 
M.~Schumacher$^1$, B.~Seitz$^1$,
F.~Smend$^1$, A.~Thomas$^2$, W.~Weihofen$^1$, and F.~Zapadtka$^1$
}

\address{
$^1$~Zweites Physikalisches Institut, Universit\"at G\"ottingen, D-37073
G\"ottingen, Germany \\
$^2$~Institut f\"ur Kernphysik, Universit\"at Mainz,
D-55099 Mainz, Germany \\
$^3$~Physikalisches Institut, Universit\"at T\"ubingen,
D-72076 T\"ubingen, Germany \\
$^4$~Department of Physics and Astronomy, University of Glasgow,
Glasgow G12 8QQ, UK \\
$^5$~B.I. Stepanov Institute of Physics, Belarussian Academy of
Sciences, 220072 Minsk, Belarus\\
$^6$~P.N. Lebedev Physical Institute, 117924 Moscow, Russia
}

\date{\today}
\maketitle
%
\begin{abstract}

Measuring Compton scattered photons and recoil neutrons in coincidence, 
quasi-free Compton scattering by the neutron has been investigated at 
MAMI (Mainz) at $\theta^{\rm lab}_\gamma = 136^\circ$ 
in an  energy range from 200 to 400 MeV. From the data a polarizability 
difference of 
$\alpha_n- \beta_n= 9.8 \pm 3.6 ({\rm stat}) {}^{+ 2.1}_{-1.1}{} ({\rm syst})
\pm 2.2 ({\rm model})$ in units of $10^{-4}{\rm fm}^3$ has been determined. 
In combination with the polarizability sum
$\alpha_n + \beta_n = 15.2 \pm 0.5$
deduced  from photo absorption data, the first precise results for the
neutron electric and magnetic 
polarizabilities,
$\alpha_n=12.5\pm 1.8 ({\rm stat}){}^{+ 1.1}_{-0.6}{} ({\rm syst})
\pm 1.1 ({\rm model})$ and
$\beta_n =2.7\mp 1.8 ({\rm stat}){}^{+0.6}_{-1.1}{} ({\rm syst})
\mp 1.1 ({\rm model})$,
are obtained.

\end{abstract}


}
The electromagnetic polarizabilities belong to the fundamental structure
constants of the nucleon.
Although attempts to measure the electromagnetic polarizabilities 
of the neutron
have a long history the results obtained up to now have remained
unsatisfactory. Since Compton scattering experiments appeared too difficult,
the first generation of investigations concentrated on the method of 
electromagnetic 
scattering of low-energy neutrons in the electric fields of heavy nuclei,
as measured in neutron transmission experiments. 
The history of these studies is summarized in Ref. \cite{aleksandr}.
The latest in a series of experiments have been carried out at Oak Ridge
\cite{Sch91} and Munich \cite{Koe95} leading to
$\alpha_n=12.6\pm1.5\pm2.0$ and $\alpha_n=0.6\pm5$, respectively,
for the electric polarizability of the neutron
in units of $10^{-4}$fm$^3$ which will be used throughout in the
following. The numbers given here have been  corrected by adding  the Schwinger
term \cite{Lvov93} $e^2 \kappa^2_{\rm n}/4M^3= 0.6$,
 containing the neutron anomalous magnetic moment $\kappa_n$ and the 
neutron mass $M$, 
 which had been omitted 
in the original evaluation of these experiments. After including
the Schwinger term  the numbers are directly
comparable with the ones defined through the Compton scattering process
\cite{Lvov93}. While the Munich  result \cite{Koe95} has a large
error,  the Oak Ridge result
\cite{Sch91} is of very  high
precision. However, this high precision has been questioned by a number of
researchers active in the field of neutron scattering
\cite{Enik97}. Their conclusion is that the Oak Ridge result \cite{Sch91}
possibly might be quoted as $7\leq \alpha_n\leq 19$.
Note that the neutron transmission experiments  do not constrain the
magnetic polarizability $\beta_n$.

The method of Compton scattering makes use of the equation 
\begin{equation}
f=f_B+\omega\omega' \alpha \, {\boldsymbol{\epsilon}}\cdot 
{\boldsymbol{\epsilon}'}
+\omega \omega' \beta \, {\bf s}\cdot {\bf s'} +{\cal O}(\omega^3),
\label{compzton}
\end{equation}
where $f_B$ is the Born amplitude, $\alpha$ the electric and $\beta$
the magnetic polarizability, $\omega$, $\omega'$ the photon
energies in the initial and final state, respectively, and  
${\boldsymbol  \epsilon}$,
${\boldsymbol \epsilon}'$ and ${\bf s}$, ${\bf s}'$  the 
directions of the corresponding electric and magnetic fields.
A pioneering experiment on Compton scattering by the neutron had been carried
out by the G\"ottingen and Mainz groups at the electron beam of MAMI A
operated at 130 MeV \cite{rose90}. This experiment followed a
proposal of Refs.\cite{levchuk94} to exploit the reaction 
$\gamma d\to \gamma np$ in
the quasi-free kinematics, though there is an evident reason why 
such an experiment is difficult at energies below  pion threshold. 
For the proton the largest portion of the polarizability-dependent cross
section in this energy region stems from the interference term between the
Born amplitude containing Thomson scattering as the largest contribution,  and 
the non-Born amplitude containing the polarizabilities. For the
neutron the Thomson amplitude vanishes so that the interference  term 
is very small and correspondingly cannot be used for the determination 
of the neutron polarizabilities. This implies that the cross section 
is rather small being about 2--3 nb/sr at 100 MeV. The way chosen to 
overcome this problem   
was to use a high flux of bremsstrahlung without
tagging\cite{rose90,levchuk94}. 
The result obtained in the experiment \cite{rose90} was
\begin{equation}
\alpha_n=10.7^{+3.3}_{-10.7}.
\label{Rose}
\end{equation}
    
This means that the experiment was successful in providing a value
for the electric polarizability and its upper limit but it did not 
permit to determine a definite lower limit. The reason for this 
deficiency is that below pion threshold the neutron Compton cross section 
is practically
independent of $\alpha_n$ if $0\alt \alpha_n \alt 10$ (see
Refs. \cite{levchuk94,wissmann98}).
In order to overcome this difficulty it was proposed to measure the neutron
polarizabilities at energies
above pion threshold with the energy range from 200 to 300 MeV
being the most promising,  since there the cross
sections are very sensitive to $\alpha_n$ \cite{levchuk94,wissmann98}
if, in addition, large scattering angles are chosen. 

A first experiment on quasi-free Compton scattering by the proton
bound in the deuteron for energies above pion threshold was carried
out at MAMI (Mainz) \cite{wissmann99}. This experiment  served as a successful test of the method
of quasi-free Compton scattering for determining $\alpha - \beta$.
Later on this method was applied to the proton and the neutron
bound in the deuteron at SAL (Saskatoon) 
\cite{kolb00}. In this experiment differential cross sections
for quasi-free Compton scattering by the proton and by the neutron were 
obtained  at a scattering angle of $\theta^{\rm lab}_\gamma = 135^\circ$
for incident  photon energies of $236-260$ MeV, which were combined to give
one data point of reasonable precision for each nucleon.
From the ratio of these two differential cross sections a most probable
value of $\alpha_n - \beta_n = 12$ was obtained with a lower
limit of $0$ and no definite upper limit. Combining their 
results \cite{kolb00} with that of Eq.\,(\ref{Rose}) \cite{rose90} the authors 
obtained the following 1-sigma 
constraints for the electromagnetic polarizabilities 
$7.6 \leq \alpha_n \leq 14.0$ and $1.2 \leq \beta_n \leq 7.6$.

It should be noted that coherent elastic (Compton) scattering by the 
deuteron provides
a further method for determining the electromagnetic polarizabilities 
of the neutron. An evaluation of first experiments \cite{Lucas94,Horn00} 
using  the theoretical model of \cite{LL95,LL00} gave the values of
$\alpha_n + \beta_n = 20 \pm 3$ and 
$\alpha_n - \beta_n = -2 \pm 3$. 
Progress in the application of this method may be expected from an 
experiment carried out at MAX-Lab \cite{Lund00} and from further
improvements of the theoretical basis for the data analysis.
Ref. \cite{LL00} also provides access to further theoretical work on Compton
scattering by the nucleon.

\begin{figure}
\epsfxsize=3.35in
\centerline{\epsfbox{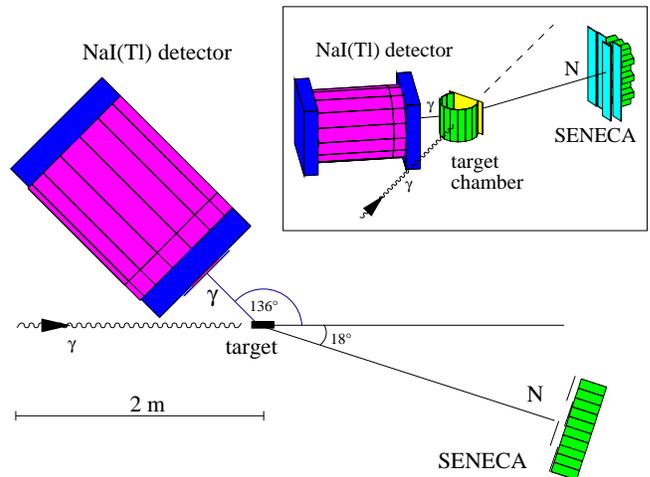}}
\caption{Experimental arrangement used for the present experiment on
Compton scattering by the proton and quasi-free Compton scattering
by the proton and the neutron. Compton scattering events were
identified through coincidences between the  Mainz 
48 cm $\O$ $\times$ 64 cm NaI(Tl) photon detector positioned at 
$\theta^{\rm lab}_\gamma = 136^\circ$ and the G\"ottingen segmented recoil 
counter SENECA positioned at $\theta^{\rm lab}_N =  18^\circ$. 
The inset shows a perspective view of this arrangement.}
\label{apparatus}
\end{figure}

In this Letter we report on first  measurements of differential cross sections
for quasi-free Compton scattering by the proton and the neutron 
covering a large  energy interval from  $E_\gamma=200$ to $400$ MeV. This
large coverage is indispensable  for determining data for the
electromagnetic polarizabilities with  good precision. 
The apparatus used is shown in Fig.~\ref{apparatus}. 
Tagged  photons produced by  the tagging facility at MAMI (Mainz) entered a 
scattering chamber,
containing a $4.6{\rm cm}\,\O \!\times 16.3 {\rm cm}$ lq.\,\,hydrogen or
lq.\,\,deuterium target in a Kapton  target cell. The 
$48{\rm cm}\,\O\!\times 64 {\rm cm}$ NaI(Tl) detector was positioned at  a 
distance of 60 cm from the target center at  a scattering angle of
$\theta^{\rm lab}_\gamma=136^\circ$ as the largest angle convenient
for an experimental set-up. A scintillation counter 
in front of the collimator is used  to identify and veto charged particles. 
As a recoil detector the G\"ottingen
SENECA detector was used, positioned at a distance of 250 cm. SENECA
was built as a neutron detector capable of pulse-shape discrimination.
It is a honeycomb structure of 30 hexagon-shaped detector cells of 15.0 cm
minimum diameter and 20.0 cm length filled with NE213 liquid scintillator.
The entrance face is covered by four plastic scintillators to
discriminate between charged and neutral particles. In the present
experiment SENECA served as the stop detector of a time-of-flight measurement,
with the start signal provided by the 
$48{\rm cm}\O\!\times 64 {\rm cm}$ NaI(Tl) detector.

Data were collected during 238 h of beam time with a deuterium target and about
35 h with a hydrogen target. The tagging efficiency was
about $55 \%$, measured several times during the runs
by means of a Pb-glass detector in the direct
photon beam, and otherwise monitored by a P2 type
ionization chamber positioned at the end of the photon beam line.
The neutron detection efficiency $\epsilon_n$ was experimentally determined
{\em in situ}
via the reaction $p(\gamma,\pi^+ n)$. The charged pion was  identified 
by the  veto counter in front of the NaI(Tl) detector and through 
the missing $\pi^+$-energy
\begin{equation} 
{\rm E}_{\pi^+}^{{\rm miss}} = {\rm E}_{\pi^+}^{{\rm calc}} - {\rm
  E}_{\pi^+}^{{\rm NaI}},
\end{equation}
where ${\rm E}_{\pi^+}^{{\rm calc}}$ is the energy of the $\pi^+$ meson
calculated from the incident photon energy and from 
the $\pi^+$ emission angle, and ${\rm E}_{\pi^+}^{{\rm NaI}}$ 
the $\pi^+$ energy measured by the NaI(Tl) detector. The result 
obtained for the neutron detection efficiency is 
$\epsilon_n = 0.180 \pm 0.014$ and proved to be in good 
agreement with previous measurements \cite{edel92}.

\begin{figure}
\epsfxsize=3.35in
\centerline{\epsfbox{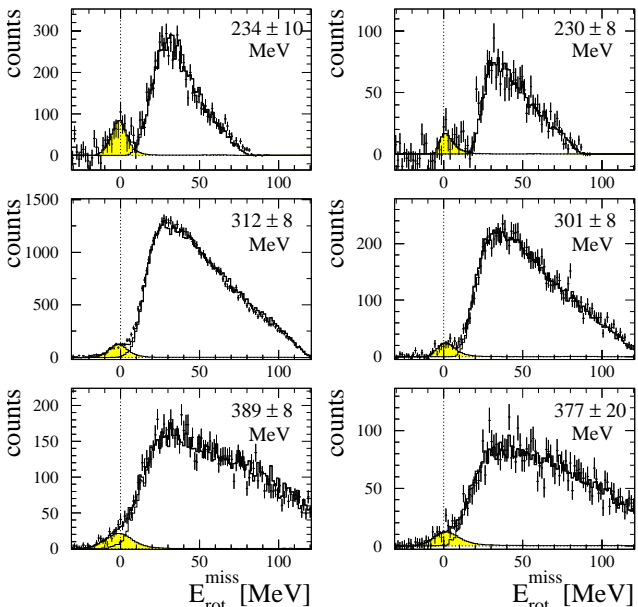}}
\caption{Spectra of photon events registered in coincidence with a 
recoil nucleon versus the missing energy of the photon 
E$^{\rm miss}_{\rm rot}$ defined in the text. 
Left panels: recoil protons identified. 
Right panels: recoil neutrons identified.
The curves are the results of a complete Monte Carlo 
simulation adjusted to the Compton scattering events (grey areas)
or to the $\pi^0$ events (white areas).} 
\label{miss-spectra}
\end{figure}

Figure \ref{miss-spectra} shows typical spectra of 
photon events measured 
in coincidence
with a recoil nucleon obtained from a  deuteron target.
In the  left panels  the recoil nucleon was identified to be a proton, 
in the right panels  a neutron. For the data analysis a two dimensional
procedure was applied with the missing nucleon  energy 
${\rm E}^{\rm miss}_N ={\rm E}^{\rm calc}_N - {\rm E}^{\rm\, SEN}_N$ 
and missing photon energy
${\rm E}^{\rm miss}_{\gamma}={\rm E}^{\rm calc}_{\gamma} -  
{\rm E}^{\rm NaI}_{\gamma}$  as 
the parameters, where
${\rm E}^{\rm\, SEN}_N$ and  ${\rm E}^{\rm NaI}_{\gamma}$ denote measured
energies and ${\rm E}^{\rm calc}_N$ and  ${\rm E}^{\rm calc}_{\gamma}$
the corresponding energies calculated assuming  a Compton event.
In this way optimal use of the separation of the two 
types of events as provided by the resolution of the apparatus has been made. 
For the spectra shown this separation of the two types of events was 
optimized by appropriately rotating the scatter 
plot of events around the origin of the 
${\rm E}^{\rm miss}_N-{\rm E}^{\rm miss}_{\gamma}$
plane before the projection of the data on the new abscissa --
${\rm E}^{\rm miss}_{\rm rot}$  -- was carried out.  
The experiment was 
accompanied by a complete Monte Carlo simulation. The curves 
shown in Fig.~\ref{miss-spectra} are the results of this Monte Carlo 
simulation 
after adjusting them to the Compton events (grey areas) and to the
$\pi^0$ events (white areas), respectively. It is apparent, 
that at the lowest 
photon energies of $\sim$  230 MeV there is a complete
separation of the two types of events. This separation remains
possible up to about 380 MeV.

For the free proton differential cross sections may be calculated
from the number of measured Compton events using a complete  Monte
Carlo simulation of the experiment to determine the detection efficiency.
For the quasi-free reaction the same procedure may be applied. However,
in this case an effective differential cross section is obtained
in this first step of data analysis which requires a second  step to take into
account the effects of binding of the nucleon in the deuteron. 
These effects of binding manifest
themselves in the Fermi momentum distribution of the nucleons. In
addition, effects due to final state interaction  of the
emitted particles and due to meson exchange currents have to
be taken into account. A detailed description of these processes
has been given in Ref. \cite{levchuk94}.
The result of the second step of  the data analysis is the triple differential
cross section in the center of the quasi-free peak of the recoil nucleon
determined
from the effective differential cross section as obtained from the number of
measured Compton events. For this determination an appropriate Monte Carlo
simulation has to be taken into account.

\begin{figure}
\epsfxsize=3.35in
\centerline{\epsfbox{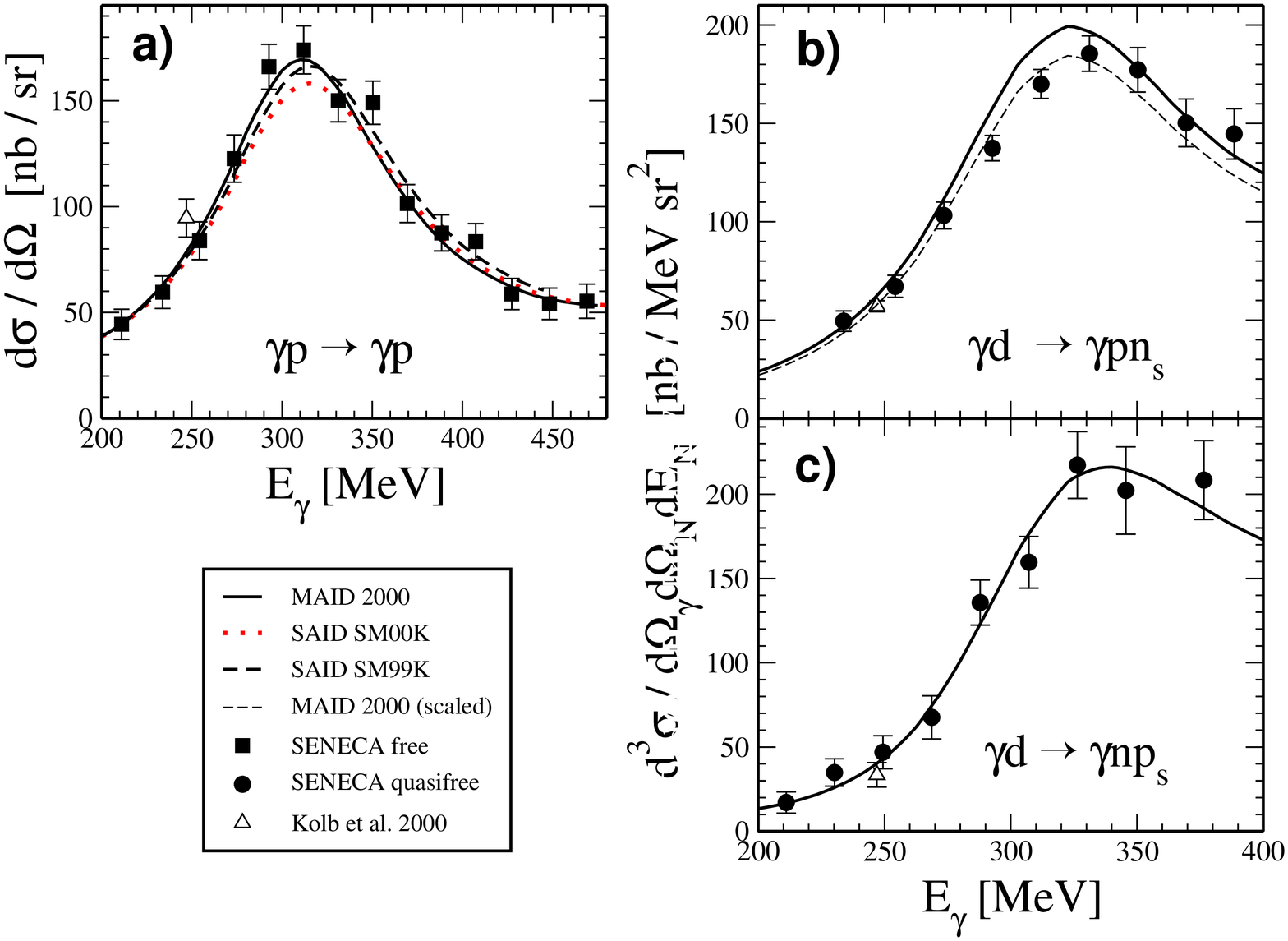}}
\caption{Differential cross sections for Compton 
scattering  at $\theta^{\rm \, lab}_\gamma$ = 136$^\circ$: 
(a) Hydrogen target with the recoil proton detected in coincidence,
(b) deuterium target with the recoil proton detected in coincidence 
and the neutron serving a spectator (s),
(c) deuterium target with the recoil neutron detected in coincidence.
Predictions: (a) Nonsubtracted dispersion theory \protect\cite{lvov97} using 
the MAID2000 (solid line), SAID-SM00K (dotted) and SAID-SM99K (dashed)
parametrizations. (b) Solid line: Same as the solid line in (a) but
calculated for the quasi-free reaction on the proton \protect\cite{levchuk94}. 
Dashed line: Same as the solid line but multiplied with a scaling factor 
of 0.93. (c) Same as the solid line in (b) but for the neutron. 
The curve is calculated for $\alpha_n - \beta_n = 9.8$. 
The SAL results \protect\cite{kolb00} are shown by triangles.
} 
\label{results}
\end{figure}

Figures \ref{results}a-c show the results of the present experiment.
The experimental data for the free proton
shown in Fig.~\ref{results}a are compared
with predictions based on the nonsubtracted dispersion theory as
described in Ref. \cite{lvov97} and thoroughly tested in 
Refs. \cite{Galler01,wolf01}. In these former experiments 
it was shown that the parametrizations SAID-SM99K \cite{SAID} and MAID2000
\cite{MAID} led to a 
good agreement with the experimental differential cross sections for Compton
scattering by the proton in the $\Delta$ resonance region, whereas
the more recent parametrization SAID-SM00K led to too small 
differential cross sections. Exactly
the same observation is  made in the present work. Therefore, the
parametrization SAID-SM00K was disregarded in the further
data analysis. Going a step further, Fig.~\ref{results}a  may be  used to find 
arguments in favor of either the MAID2000  or the SAID-SM99K parametrization.
Though the differences are small, there is a slight  preference for
the MAID2000 pameterization which is seen in Fig.~\ref{results}a and 
reflected by the  $\chi^2$ values. Therefore, we decided 
to base the further evaluation
on the MAID2000 parametrization and to use the SAID-SM99K parametrization
only for getting an estimate for the model error connected with imperfections
of the photomeson amplitudes.

In Fig.~\ref{results}b a test of  the method of quasi-free Compton 
scattering  is carried out. The data shown are triple differential cross 
sections for quasi-free Compton scattering by the proton in the center of the
quasi-free peak with their
statistical errors. The systematic errors amount to $ \pm 4.4 \%$.
The solid curve in this figure  shows triple differential
cross sections for quasi-free Compton scattering by the proton
in the center of the quasi-free peak. This theoretical prediction has been
obtained in the model of Ref. \cite{levchuk94} on
the basis of the MAID2000 parametrization. 
The $NN$-interaction entering into this model has been taken from the
CD-Bonn  potential  \cite{Machl01}. For comparison also the
separable representation of the Paris
potential \cite{PEST} has been applied, leading to essentially no difference.
The overall agreement between experiment and prediction as given by the
solid curve may be considered as
satisfactory, although there is some deviation visible in the energy range   
between 270 and 300 MeV. At present we do not have an explanation
for this residual deviation which, therefore, could not be eliminated 
by means of  a correction. Consequently, 
we have to treat it as a possible source of uncertainty which has to be 
taken into account through a further contribution to the model error.
In order to get a quantitative result for this additional model error,
the prediction shown in Fig.~\ref{results}b as a solid curve
has been scaled down by a factor of 0.93 to give the dashed curve. 
Through this procedure
we arrive at a modified set of photomeson amplitudes which may be used 
in the further analysis to estimate the additional model error connected
with a possible imperfection of the theory of the quasi-free
Compton scattering.

Figure \ref{results}c shows triple differential cross sections for the neutron
in the center of the
quasi-free peak compared with predictions obtained in the
model of Ref. \cite{levchuk94} on the basis of the MAID2000 parametrization.
The difference between the methods of evaluation  in 
Fig.~\ref{results}b and Fig.~\ref{results}c is that for the proton  
the parameter
$\alpha_p - \beta_p= 10.5\pm 1.5$ is fixed through additional experiments
\cite{Olmos01}
whereas for the neutron $\alpha_n - \beta_n$ is a free parameter
which has to be   determined through fits to the experimental data 
using a $\chi^2$ procedure. The result obtained is  $\alpha_n - \beta_n$ = 9.8.
The errors of this result are as follows.
The statistical error from the $\chi^2$ procedure is $\pm$3.6.
The systematic error of the neutron triple differential cross sections
amounts to $\pm 9\%$, with  the detection efficiency $\epsilon_n$
of the neutrons contributing $\pm 8\%$, the number of target nuclei 
per cm$^2$ contributing $\pm 2\%$, the  uncertainties caused by cuts 
in the spectra and by the Monte Carlo simulations contributing 
$\pm 3\%$ and the tagging 
efficiency contributing $\pm 2.5\%$. For $\alpha_n - \beta_n$ this 
leads to a combined probable  systematic error of ${}^{+2.1}_{-1.1}$. The model
error due to imperfections of the parametrization of photomeson
amplitudes was estimated from a comparison of results
obtained with the MAID2000 and SAID-SM99K parametrizations, respectively. 
The result obtained for $\alpha_n - \beta_n$ is $\pm 2.0$. 
The errors due to different parametrizations of the $NN$-interaction 
is found to be about $\pm 0.2$. 
The determination
of the model error due to possible imperfections of the theory
of quasi-free Compton scattering has been discussed above in connection
with Fig.~\ref{results}b and amounts to $\pm 0.8$.

Taking all these errors into account we arrive at our final result
\begin{equation}
\alpha_n- \beta_n= 9.8 \pm 3.6 ({\rm stat}) {}^{+ 2.1}_{-1.1}{} ({\rm syst})
\pm 2.2 ({\rm model}).
\end{equation}
Combining it with 

$\alpha_n + \beta_n = 15.2 \pm 0.5$ \cite{LL00} we obtain
\begin{eqnarray}
\alpha_n&=12.5\pm 1.8 ({\rm stat}){}^{+ 1.1}_{-0.6}{} ({\rm syst})
\pm 1.1 ({\rm model}),\\ 
\beta_n &=~~2.7\mp 1.8 ({\rm stat}){}^{+0.6}_{-1.1}{} ({\rm syst})
\mp 1.1 ({\rm model}).
\end{eqnarray}

It is of interest to compare the present result obtained for the
neutron with the corresponding result for the proton. Combining the 
global averages of the electric and magnetic polarizabilities 
determined  in Ref. \cite{Olmos01} with the value for the
sum of polarizabilities $\alpha_p + \beta_p = 14.0 \pm 0.3$
obtained in Ref. \cite{LL00} we arrive at
\begin{eqnarray}
\alpha_p& =12.2\pm 0.3 ({\rm stat})\mp 0.4 ({\rm syst})\pm 0.3 ({\rm model}),\\
\beta_p& =~~1.8\pm 0.4 ({\rm stat})\pm 0.4 ({\rm syst})\pm 0.4 ({\rm model}). 
\end{eqnarray}
The comparison shows  that there is no significant isovector component
in the electromagnetic polarizabilities. Nevertheless, there is a slight 
tendency suggesting that the magnetic polarizability of the neutron may be
larger than that of the proton.


{\small 
One of the authors (M.I.L.) highly appreciates the hospitality of
II. Physikalisches Institut der Universit\"at G\"ottingen where part of his
theoretical work was done.
This work was supported by Deutsche Forschungsgemeinschaft 
(SFB 201 and  SFB 433 Mainz), and by Schwer\-punktprogramm (1034) through 
the contracts DFG-Wi1198, DFG-Schu222, and through the German Russian
exchange program 436 RUS 113/510.
}


\end{document}